\documentclass[12pt]{iopart}

\usepackage{t1enc}
\usepackage[utf8]{inputenc}

\usepackage{eucal}

\usepackage[usenames,dvipsnames]{xcolor}
\usepackage{hyperref}
\usepackage{graphicx}
\usepackage{amssymb, amsfonts}

\newcommand{\dd}{\mathrm{d}}

\newcommand{\cplxi}{\mathrm{i}}
\newcommand{\expe}{\mathrm{e}}

\begin{document}

\title[Quantum Particle Motion in Absorbing Harmonic Trap]{Quantum Particle Motion in Absorbing Harmonic Trap}

\author{B G M\'{a}rkus$^1$, F M\'{a}rkus$^{1,\dagger}$}
\address{$^1$ Budapest University of Technology and Economics, Department of Physics, \\ H-1521 Budapest Budafoki út 8., Hungary}
\ead{$^{\dagger}$markus@phy.bme.hu; f.markus@eik.bme.hu}

\begin{abstract}
The motivation of this work is to get an additional insight into the irreversible energy dissipation on the quantum level. The presented examination procedure is based on the Feynman path integral method that is applied and widened towards the calculation of the kernel of a quantum mechanical damped oscillator. Here, it is shown that the energy loss of the oscillator can be generated by the introduced harmonic complex potential. The related damped wave function, however, does not pertain to the probability meaning as it is usual in the case of complex absorbing potentials. This decrease of the wave function is evaluated, moreover, the energy dissipation and the measure the irreversibility are expressed.
\end{abstract}


\vspace{2pc}


\maketitle

\section{Introduction}

\subsection{Preliminaries}

The idea of introducing complex absorbing potentials is originated from the description of scattering processes [1-7]. In these studies it is always strongly emphasized that these kind of potentials relate to a kind of dissipative behavior of the considered phenomenon. Usually, it is useful to treat some simple physical system to achieve the deeper meaning from the behavior under the complex potentials in the dissipation and the irreversibility. The results on a simple system can be used for further investigation; for example, this may happen during photon absorption process (e.g., infra-red spectroscopy \cite{kuzmany2009}) where the effect of solvent could be treated in this frame. It can be studied that in an absorption-emission process how the solvent can also drain energy from the molecular oscillations. On the other hand, during the discovery of Bose--Einstein condensates (BEC) the thermal cloud (non-condensed particles) create a damping effect \cite{Zaremba1999,Jackson2002}. It is known that BEC also behaves as an oscillator (in a parabolic trap), the calculations can also be useful for that phenomenon of physics. Open systems might also be studied with the current procedure. That is why we make example calculations on the damped quantum oscillator first. 

\subsection{Schr\"odinger equation with complex absorbing potential}

The Schr\"odinger equation can be generalized [1-7,11] by applying a complex potential in the form $V_r(x) + \cplxi V_{c}(x)$, where the introduced $V_r(x)$ and $V_{c}(x)$ are real-valued, thus
\begin{equation}  \label{eq:sch4_2}
\frac{\hslash}{\cplxi} \frac{ \partial \Psi }{\partial t} - \frac{\hslash^2}{2m} \nabla^2 \Psi + \left(V_{r} + \cplxi V_{c}\right) \Psi = 0,
\end{equation}
and parallel its complex conjugate is
\begin{equation}  \label{eq:sch5}
-\frac{\hslash}{\cplxi} \frac{ \partial \Psi^{\ast}}{\partial t} - \frac{\hslash^2}{2m} \nabla^2 \Psi^{\ast} + \left(V_{r} - \cplxi V_{c}\right) \Psi^{\ast} = 0.
\end{equation}
Due to the sign change at the complex potential term, instead of the continuity equation, we obtain a balance equation for the expression $\Psi^{\ast} \Psi$:
\begin{equation}  \label{eq:sch7}
\frac{\partial \left(\Psi^{\ast} \Psi \right)}{\partial t} + \frac{\cplxi \hslash}{2m} \nabla \left(\Psi \nabla \Psi^{\ast} - \Psi^{\ast} \nabla \Psi\right)
- \frac{2}{\hslash}V_{c} \Psi^{\ast}\Psi = 0.
\end{equation}
The third term is a source density thus the expression
\begin{equation}
\rho = \Psi^{\ast} \Psi
\end{equation}
can only have a probability meaning if $V_c =0$ (in the conventional manner). Since, Eq. (\ref{eq:sch7}) is a balance equation, depending on the sign of the source, presently $V_{c}$, the integral $\int_{V} \Psi^{\ast} \Psi \, \dd V$ in the all space may decrease or increase in time. It sounds physically acceptable that if $\int_{V} \Psi^{\ast} \Psi \, \dd V$ changes in time, the {\it decrease} is the relevant choice to describe losses. The case of decrease is limited from below by the $0$, however, the case of increase is not necessarily limited from above. Thus, from this physical reason the complex potential $V_{c}$ must be negative which will describe the dissipation or loss in the system. The probability meaning can also be generalized to negative complex potential cases with the meaning that in some measurements the particle seems to be missing. Thus $\rho$ represents the remaining fraction of the initial ensemble.

\subsection{The propagator of the harmonic oscillator}

\begin{figure}[h!]
\centerline{
\includegraphics[width=0.5\columnwidth]{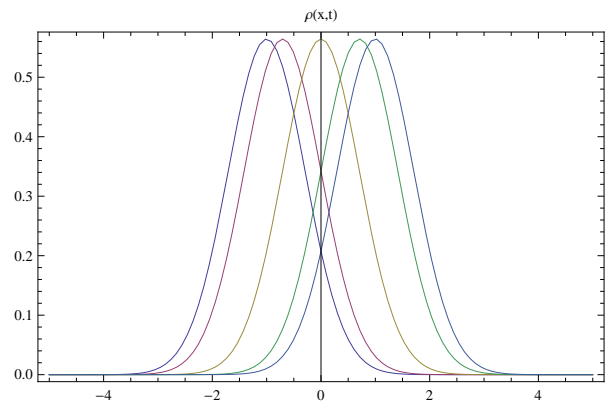}}
\caption{The movement of the probability distribution $\rho(x,t)$ of the undamped oscillator in the first half period. The calculation is evaluated by the parameter set: $m = 1$; $\hslash = 1$; $\omega = 1$. The peak of the initial distribution is at $x = y_0 = -1$. The curves pertain to: {\color{blue}$t = 0$} (blue); {\color{purple}$t = T/8$} (purple); {\color{brown}$t = T/4$} (brown); {\color{green}$t = 3T/8$} (light green) and {\color{RoyalBlue}$t = T/2$} (light blue).}  \label{const_shape_rho_in_half_period}
\end{figure}

First focus on the case of the one-dimensional quantum mechanical harmonic oscillator whose kernel and the time evolution of its wave function can be solved analytically by the Feynman path integral method \cite{Dittrich2001}. For this reason the start of the examination from the Lagrangian of the classical harmonic oscillator seems to be relevant. That can be written as it is usual in the formulation of
\begin{equation}
L = K - V = \frac{1}{2}m{\dot{x}}^2 - \frac{1}{2} m {\omega}^2 x^2.
\end{equation}
The related action $S$ during the considered motion between the $t_1 \rightarrow t_2$ time interval can be calculated by the evaluation of the integral $S = \int_{t_1}^{t_2} L \, \dd t$. Through the detailed calculations \cite{Dittrich2001} we formulate the action as
\begin{equation}  \label{action_real}
S = \frac{m\omega}{2\sin (\omega t)} \left[ \left(x_2^2+x_1^2\right) \cos (\omega t) - 2x_1 x_2 \right],
\end{equation}
where the elapsed time is denoted by $t = t_2 - t_1$; the $x_1$ and $x_2$ are the initial and final space coordinates at $t_1$ and $t_2$, respectively. In the sense of the Feynman path integral method  the propagator (kernel) can be formulated as
\begin{equation}
K(x_2,t,x_1,0) = A(t)\expe^{\frac{\cplxi}{\hslash}S},
\end{equation}
where $A(t)$ is a time dependent amplitude factor. After the normalization procedure the propagator of the harmonic oscillator can be expressed
\begin{eqnarray}
K(x_2,t,x_1,0) = \sqrt{\frac{m\omega}{2 \pi \cplxi \hslash \sin (\omega t)}} \nonumber \\ \times \exp \left\{ \frac{\cplxi}{2\hslash} \frac{m\omega}{\sin (\omega t)} \left[ (x_2^2+x_1^2) \cos (\omega t) - 2x_1 x_2 \right] \right\}.
\end{eqnarray}
Here, there are problematic points at the caustics, i.e., when $t_n = n\pi/ \omega$. It is necessary to avoid this mathematical conflict (singularities both in the square root and in the exponential). Therefore, the correct propagator for the harmonic oscillator needs to be recalculated. The final result can be expressed by the Feynman--Soriau formula
\begin{eqnarray}\label{eq:fs}
K(x_2,t,x_1,0) = \exp \left\{ -\cplxi\frac{\pi}{2} \left( \frac{1}{2} + \left\lfloor \frac{\omega t}{\pi} \right\rfloor  \right) \right\}
\sqrt{\frac{m\omega}{2 \pi \cplxi \hslash \sin (\omega t) }} \nonumber \\ \times \exp \left\{ \frac{\cplxi}{2\hslash} \frac{m\omega} {\sin (\omega t)}
\left[ (x_2^2+x_1^2) \cos (\omega t) - 2x_1 x_2 \right] \right\},
\end{eqnarray}
where the first multiplier as a phase factor resolves the problem. (The $\lfloor \cdots \rfloor$ denotes the floor operation.) The time evolution of the wave function from the initial space-time $(y,0)$ to $(x,t)$ is governed by the propagator as
\begin{equation}
{\Psi}(x,t) = \int\limits_{-\infty}^{\infty} K(x,t,y,0) \Psi_{0}(y,0) \, \dd y .
\end{equation}
The oscillation starts from a normalized Gaussian initial wave function
\begin{equation}  \label{initial_Gaussian}
\Psi_{0}(y,0) = \sqrt[4]{\frac{ m\omega}{\pi\hslash}} \exp \left( - \frac{m\omega}{2\hslash}(y-y_0)^2 \right)
\end{equation}
with its centre position $y_0$, the ${\Psi}(x,t)$, and then the $\rho(x,t)$ distribution can be evaluated. A time series of $\rho(x,t)$ with parameters set as $m = 1$, $\hslash = 1$ and $\omega = 1$ is shown in Fig. \ref{const_shape_rho_in_half_period} in which the evolution can be easily followed. The time period is $T = {2\pi}/{\omega}$.  

%\begin{figure}[h!]
%\centerline{
%\includegraphics[width=0.5\columnwidth]{const_shape_rho_in_half_period.eps}}
%\caption{The movement of the probability distribution $\rho(x,t)$ of the undamped oscillator in the first half period. The calculation is evaluated by the parameter set: $m = 1$; $\hslash = 1$; $\omega = 1$. The peak of the initial distribution is at $x = y_0 = -1$. The curves pertain to: {\color{blue}$t = 0$} (blue); {\color{purple}$t = T/8$} (purple); {\color{brown}$t = T/4$} (brown); {\color{green}$t = 3T/8$} (light green) and {\color{RoyalBlue}$t = T/2$} (light blue).}  \label{const_shape_rho_in_half_period}
%\end{figure} 

\begin{figure}[h!]
\centerline{
\includegraphics[width=0.5\columnwidth]{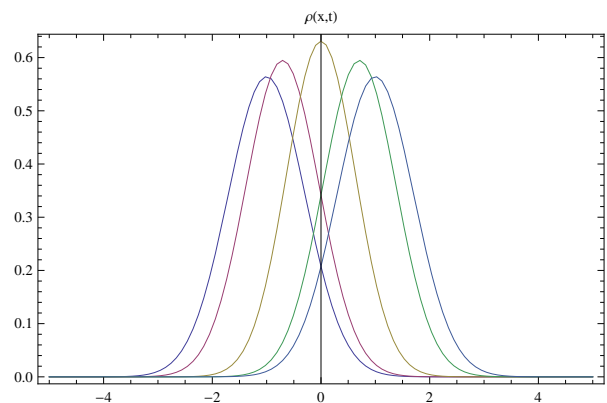}}
\caption{The $\gamma$-governed shape change during the movement of the probability distribution $\rho(x,t)$ of the undamped oscillator in the first half period. The calculation is evaluated by the parameter set: $m = 1$; $\hslash = 1$; $\omega = 1$ and $\gamma = 0.8$. The peak of the initial distribution is at $x = y_0 = -1$. The curves pertain to: {\color{blue}$t = 0$} (blue); {\color{purple}$t = T/8$} (purple); {\color{brown}$t = T/4$} (brown); {\color{green}$t = 3T/8$} (light green) and {\color{RoyalBlue}$t = T/2$} (light blue).}  \label{shape_change_rho_in_half_period}
\end{figure}

A detailed study of the oscillator wave packet motion was elaborated by Naqvi and Waldenstrøm \cite{Razi2000}. It is shown that by introducing a $\gamma \neq 1$ parameter, the width of the wave packet also "oscillates", i.e., the Gaussian shape periodically changes in time described by the expression
\begin{equation}
| \Psi (x,t) |^2 = \frac{1}{\sigma_{x}(t)\sqrt{2}} \exp{ \left\{ -\frac{\left[x - x_{0} \cos(\omega t) \right]^2}{2\sigma_{x}^{2}(x,t)} \right\}},
\end{equation} 
where
\begin{eqnarray}
\sigma_{x}^{2}(x,t) = \sigma_{x}^{2}(0) \left[ \cos^2(\omega t) + {\gamma}^2 \sin^2(\omega t) \right], \\ \sigma_{x}^{2}(0) = \frac{\hslash}{2{\gamma}m{\omega}}.   
\end{eqnarray}
This wave packet motion can be reproduced applying the kernel
\begin{eqnarray}\label{eq:fs2}
K(x_2,t,x_1,0) = \exp \left\{ -\cplxi\frac{\pi}{2} \left( \frac{1}{2} + \left\lfloor \frac{\omega t}{\pi} \right\rfloor  \right) \right\}
\sqrt{\frac{m\omega}{2 \pi \cplxi \hslash \sqrt{\gamma} \sin (\omega t) }} \nonumber \\ \times \exp \left\{ \frac{\cplxi}{2\hslash} \frac{m\omega} {\sqrt{\gamma} \sin (\omega t)}
\left[ (x_2^2+x_1^2) \cos (\omega t) - 2x_1 x_2 \right] \right\}
\end{eqnarray}
with the initial condition given by Eq. (\ref{initial_Gaussian}) as Fig. \ref{shape_change_rho_in_half_period} shows.

%\begin{figure}[h!]
%\centerline{
%\includegraphics[width=0.5\columnwidth]{shape_change_rho_in_half_period.eps}}
%\caption{The $\gamma$-governed shape change during the movement of the probability distribution $\rho(x,t)$ of the undamped oscillator in the first half period. The calculation is evaluated by the parameter set: $m = 1$; $\hslash = 1$; $\omega = 1$ and $\gamma = 0.8$. The peak of the initial distribution is at $x = y_0 = -1$. The curves pertain to: {\color{blue}$t = 0$} (blue); {\color{purple}$t = T/8$} (purple); {\color{brown}$t = T/4$} (brown); {\color{green}$t = 3T/8$} (light green) and {\color{RoyalBlue}$t = T/2$} (light blue).}  \label{shape_change_rho_in_half_period}
%\end{figure} 

\noindent This propagator formulation will be generalized to the complex potential case to achieve the description of the damped quantum oscillator.

\section{Damped quantum oscillator with complex absorbing harmonic potential}

\begin{figure}[h!]
\centerline{
\includegraphics[width=0.4\columnwidth,angle=270]{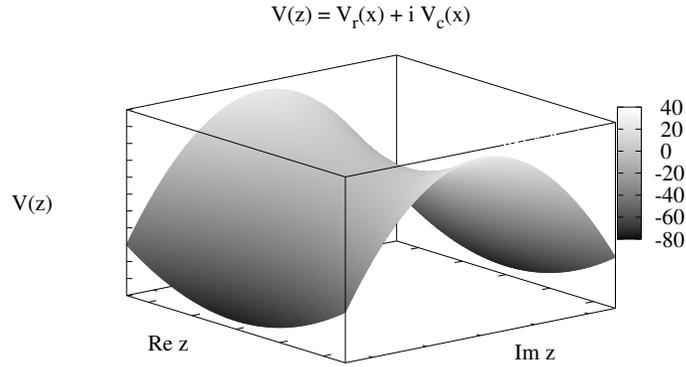}}
\caption{A demonstrative 3D plot of the harmonic complex potential $V(x)$ in Eq. (\ref{nagative_imaginary_potential}) with the negative imaginary part: $V_{c}(x) = - \frac{1}{2} m {\omega_c^2} x^2 = - \frac{1}{2} m {\omega_r^2} {\hat\omega}^2 x^2 < 0$. The scales of the axes are in arbitrary units.}  \label{complex_potential}
\end{figure}

Now, we can turn to our present example in which the complex harmonic potential --- shown in Fig. \ref{complex_potential} --- is introduced in the following form
\begin{eqnarray}  \label{nagative_imaginary_potential}
V(x) &= \frac{1}{2} m \left( {\omega_r^2} - \cplxi{\omega_c^2} \right) x^2 = \frac{1}{2} m {\omega_r^2} \left( 1  - \cplxi \left( \frac{\omega_c}{\omega_r} \right)^2 \right) x^2  \nonumber \\
&= \frac{1}{2} m {\omega_r^2} \left( 1  - \cplxi \hat\omega^2 \right) x^2 .
\end{eqnarray}
%
%\begin{figure}[h!]
%\centerline{
%\includegraphics[width=0.4\columnwidth,angle=270]{neg}}
%\caption{A demonstrative 3D plot of the harmonic complex potential $V(x)$ in Eq. (\ref{nagative_imaginary_potential}) with %the negative imaginary part: $V_{c}(x) = - \frac{1}{2} m {\omega_c^2} x^2 = - \frac{1}{2} m {\omega_r^2} {\hat\omega}^2 x^2 < 0$. The scales of the axes are in arbitrary units.}  \label{complex_potential}
%\end{figure}

\noindent As it can be seen, both potential terms are quadratic in $x$. Here, the real valued angular frequencies $\omega_r$ and $\omega_c$ pertain to the real and the imaginary parts. To shorten the formulas we introduce the abbreviation $\hat\omega = \omega_c/\omega_r$. Accepting the above idea of the potential $V(x)$, the complex Lagrangian can be expressed as
\begin{equation}
L = \frac{1}{2}m{\dot{x}}^2 - \frac{1}{2} m {\omega_r^2} \left( 1  - \cplxi \hat\omega^2 \right) x^2.
\end{equation}
The action of the present problem can be obtained by the direct substitution of the complex frequency $\omega$ in Eq. (\ref{action_real})
\begin{equation}  \label{complex_omega}
\omega \longrightarrow \omega = \sqrt{{\omega_r^2} - \cplxi{\omega_c^2}} = \omega_r \sqrt{1+ \hat\omega^4} \,\, \expe^{\cplxi{\varphi}/{2}},
\end{equation}
keeping the physically correct signs. Here, the phase $\varphi$ can be expressed by
\begin{equation}
\varphi = \arctan \left(-\hat\omega^2\right) \,\, \leq 0.
\end{equation}
It is easy to check that the real part of $\omega$ in Eq. (\ref{complex_omega}) is exactly the frequency of the undamped harmonic oscillator
\begin{equation}
{\omega}_{r} = \textrm{Re}(\omega) = \omega_r \sqrt{1+ \hat\omega^4} \cos\frac{\varphi}{2}.
\end{equation}
While the imaginary part of $\omega$ relates to the factor ${\omega}_{c} = \beta$ --- that expresses the friction ($\sim {\beta}v$) in the classical mechanical systems ---, i.e.
\begin{equation}
{\omega}_{c} = \beta = \textrm{Im}(\omega) = \omega_r \sqrt{1+ \hat\omega^4} \sin\frac{\varphi}{2}.
\end{equation}
We note that the sign of the ${\omega}_{c} \, (= \beta)$ is negative. Due to the imaginary term in the Lagrangian the action $S_{\mathrm{im}}$ becomes also complex
\begin{eqnarray}  \label{action_imaginary}
S_{\mathrm{im}} = \frac{m\omega_r \sqrt{1+ \hat\omega^4} \, \expe^{\cplxi {\varphi}/{2}}}{2\sin (\omega_r \sqrt{1+ \hat\omega^4} \, \expe^{\cplxi {\varphi}/{2}} t)} \nonumber \\ \times \left[ \left(x_2^2+x_1^2\right) \cos \left(\omega_r \sqrt{1+ \hat\omega^4} \, \expe^{\cplxi {\varphi}/{2}} t\right) - 2x_1 x_2 \right].
\end{eqnarray}
For further calculations it is useful to write the sine in the denominator as
\begin{equation}
\sin \left(\omega_r \sqrt{1+ \hat\omega^4} \, \expe^{\cplxi {\varphi}/{2}} t\right) = \sin \left(\omega_r t\right) \cosh\left(\omega_c T\right) +
\cos \left(\omega_r t\right) \sinh\left(\omega_c T\right),
\end{equation}
while the cosine factor in the coordinate dependent part is
\begin{equation}
\cos \left(\omega_r \sqrt{1+ \hat\omega^4} \, \expe^{\cplxi {\varphi}/{2}} t\right) = \cos \left(\omega_r t\right) \cosh\left(\omega_c t\right) -
\sin \left(\omega_r t\right) \sinh\left(\omega_c t\right).
\end{equation}
Summarizing the previous considerations, the propagator of the damped oscillator can be expressed as 
\begin{eqnarray}
K(x_2,t,x_1,0) = \mathcal{N}\expe^{\frac{\cplxi}{\hslash}S_{\mathrm{im}}} \nonumber \\ =
\mathcal{N}\exp \left\{ \frac{\cplxi}{\hslash} \frac{m\omega_r \sqrt{1+ \hat\omega^4} \, 
\expe^{\cplxi {\varphi}/{2}}} {2 \left( \sin \left(\omega_r t\right) \cosh\left(\omega_c T\right) +
\cos \left(\omega_r t\right) \sinh\left(\omega_c T\right) \right)} \right. \nonumber \\
\times \left. \left[ \left(x_2^2+x_1^2\right) \left( \cos \left(\omega_r t\right) \cosh\left(\omega_c t\right) -
\sin \left(\omega_r t\right) \sinh\left(\omega_c t\right) \right) - 2x_1 x_2 \right] \right\}.
\end{eqnarray}
where $\mathcal{N}$ is the normalization factor. However, due to the dissipation, $\mathcal{N}$ must be defined in the initial time $t = 0$ by the following equation in one dimension:
\begin{equation}
1 := \left. \int\limits_{-\infty}^{\infty} \Psi^{\ast} \Psi \, \dd x \right|_{t=0}.
\end{equation}
From the physical considerations it seems obvious that this factor decreases in time. Naturally, it must also involve the undamped case, i.e., when $\hat\omega =0$. After the evaluation of the detailed calculations we obtain the expression of the kernel 
\begin{eqnarray}
K(x,t,y,0) = \exp \left\{ -\cplxi \frac{\pi}{2} \left( \frac{1}{2} + \left\lfloor \frac{\sqrt{\omega_{r}^{2} - \omega_{c}^{2}} t}{\pi} \right\rfloor \right) \right\} \exp \left\{ \frac{\omega_{c} t}{2\gamma} \right\} \nonumber \\ \times
\sqrt{\frac{m\omega_{r} \sqrt{\gamma}}{2 \pi \cplxi \hslash
\left\{ \gamma \sin \left(\omega_r t\right) \cosh\left(\omega_c t\right) + \cos \left(\omega_r t\right) \sinh\left(\omega_c t\right) \right\}}} \nonumber \\
\times \exp \left( \cplxi \zeta_{r}(t)x^2 + \cplxi \zeta_{r}(t)y^2 - 2 \cplxi \tilde{\zeta}_r(t) xy \right) \nonumber \\ \times
\exp \left( - \zeta_{c}(t)x^2 - \zeta_{c}(t)y^2 + 2 \tilde{\zeta}_c(t) xy \right),
\end{eqnarray}
where the introduced functions are
\[
\zeta_{r}(t) = \frac{\sqrt{\gamma} m \omega_{r}}{2\hslash} \frac{\cos{\omega_{r}t} \cosh{\omega_c}t - \gamma \sin{\omega_{r}t} \sinh{\omega_c}t}{\gamma  \sin{\omega_{r}t} \cosh{\omega_c}t + \cos{\omega_{r}t} \sinh{\omega_c}t},
\]

\[
\tilde{\zeta}_{r}(t) = \frac{\sqrt{\gamma} m \omega_{r}}{2\hslash} \frac{1}{\gamma  \sin{\omega_{r}t} \cosh{\omega_c}t + \cos{\omega_{r}t} \sinh{\omega_c}t},
\]

\[
\zeta_{c}(t) = \frac{\sqrt{\gamma} m \omega_{c}}{2\hslash} \frac{\cos{\omega_{r}t} \cosh{\omega_c}t - \gamma \sin{\omega_{r}t} \sinh{\omega_c}t}{\gamma  \sin{\omega_{r}t} \cosh{\omega_c}t + \cos{\omega_{r}t} \sinh{\omega_c}t},
\]

\[
\tilde{\zeta}_c(t) = \frac{\sqrt{\gamma} m \omega_{c}}{2\hslash} \frac{1}{\gamma  \sin{\omega_{r}t} \cosh{\omega_c}t + \cos{\omega_{r}t} \sinh{\omega_c}t}.
\]
The initial Gaussian distribution is written in the parameter fitted form 
\begin{equation}  \label{gamma_omega_r_initial_Gaussian}
\Psi_{\mathrm{init}}(y,0) = \sqrt[4]{\frac{m\omega_r}{\pi\hslash}} \exp \left( - \frac{m\omega_r}{2\hslash}(y-y_0)^2 \right).
\end{equation}

\section{Results and discussion}

\begin{figure}[h!]
\centerline{
\includegraphics[width=0.5\columnwidth]{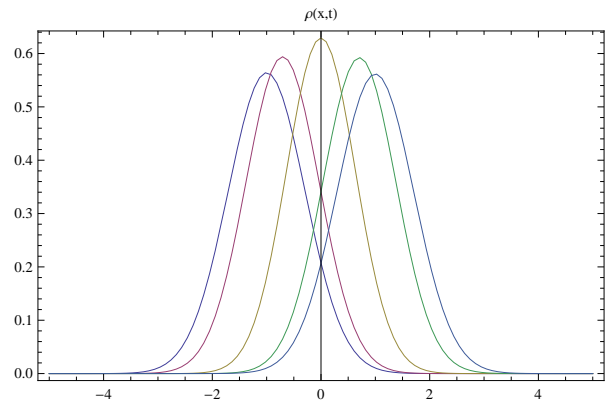}}
\caption{The change of the shape of distribution $\rho(x,t)$ in the first half period. The calculation is evaluated by the parameter set: $m = 1$; $\hslash = 1$; $\omega_r = 1$, $\hat\omega = 0.05$ and $\gamma = 0.8$. The initial position is at $x = y_0 = -1$. The {\color{blue}blue} curve on the left hand side pertains to the initial condition. The equidistant time steps between the {\color{purple}purple} -- {\color{brown}brown} -- {\color{green}light green} -- {\color{RoyalBlue}light blue} curves are ${\pi}/({4\sqrt{\omega_r^2 - \omega_c^2}})$.}  \label{rho_in_half_period}
\end{figure}

The calculations are evaluated by numerical programming. As Fig. \ref{rho_in_half_period} shows the width of the distributions depends on the parameter $\gamma$ as is in the case of undamped oscillator \cite{Razi2000}. In the present case of $\gamma = 0.8 < 1$, the $\rho(x,t)$ is thinner when the particle is going through the equilibrium position. The real part of the potential is $\omega_r = 1$ and the complex part is $\hat\omega = 0.05$. For the sake of simplicity let the parameters be $m = 1$; $\hslash = 1$. The initial position is at $x = y_0 = -1$.  

%\begin{figure}[h!]
%\centerline{
%\includegraphics[width=0.5\columnwidth]{rho_in_half_period.eps}}
%\caption{The change of the shape of distribution $\rho(x,t)$ in the first half period. The calculation is evaluated by the parameter set: $m = 1$; $\hslash = 1$; $\omega_r = 1$, $\hat\omega = 0.05$ and $\gamma = 0.8$. The initial position is at $x = y_0 = -1$. The {\color{blue}blue} curve on the left hand side pertains to the initial condition. The equidistant time steps between the {\color{purple}purple} -- {\color{brown}brown} -- {\color{green}light green} -- {\color{RoyalBlue}light blue} curves are ${\pi}/({4\sqrt{\omega_r^2 - \omega_c^2}})$.}  \label{rho_in_half_period}
%\end{figure} 

\begin{figure}[h!]
\centerline{
\includegraphics[width=0.5\columnwidth]{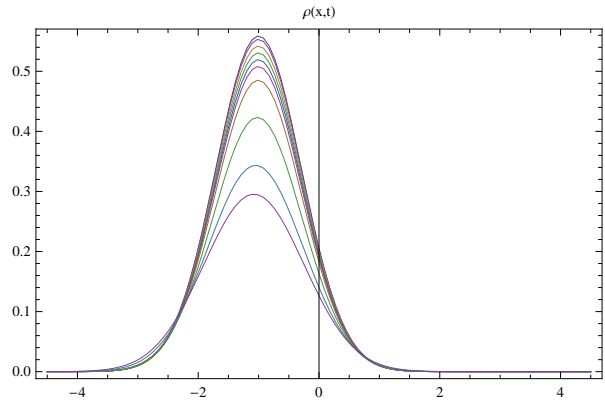}}
\caption{The decrease of distribution $\rho(x,t)$ in the (initial) position $x = -1$ after the time periods: $T, 2T, 4T, 6T, 8T, 10T, 14T, 25T, 40T, 50T$. The calculation is evaluated by the parameter set: $m = 1$; $\hslash = 1$; $\omega_r = 1$, $\hat\omega = 0.05$ and $\gamma = 0.8$. (The minor asymmetric movements come from the numerical calculation errors.)}  \label{damping_several_periods}
\end{figure}

\noindent The dissipation can be recognized adequately within a longer time interval. This damping effect can be seen in Fig. \ref{damping_several_periods}, in which the series of distributions (in different $n \cdot T$ time periods /not each/; $n = 0, 1, 2,\dots, N$) are plotted at the given position $x_0 = -1$.  

%\begin{figure}[h!]
%\centerline{
%\includegraphics[width=0.5\columnwidth]{damping_several_periods.eps}}
%\caption{The decrease of distribution $\rho(x,t)$ in the (initial) position $x = -1$ after the time periods: $T, 2T, 4T, 6T, 8T, 10T, 14T, 25T, 40T, 50T$. The calculation is evaluated by the parameter set: $m = 1$; $\hslash = 1$; $\omega_r = 1$, $\hat\omega = 0.05$ and $\gamma = 0.8$. (The minor asymmetric movements come from the numerical calculation /computational/ errors.)}  \label{damping_several_periods}
%\end{figure}

\begin{figure}[h!]
\centerline{
\includegraphics[width=0.5\columnwidth]{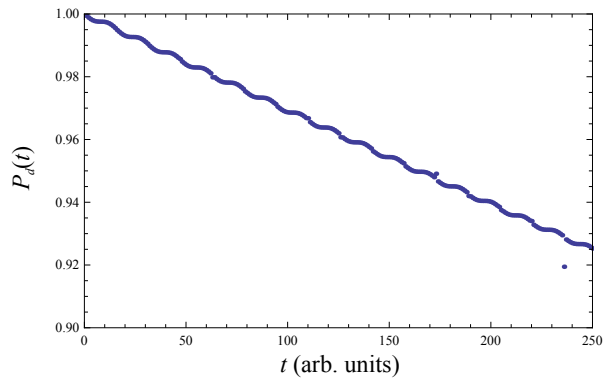}}
\caption{The decrease of ${\it P}_{d}(t) = \int_{-\infty}^{\infty} \Psi^{\ast}(x,t)\Psi(x,t) \, \dd x \leq 1$ in time is showing the dissipation during the process. The out-of-line points come from numerical errors. The calculation is evaluated by the parameter set: $m = 1$; $\hslash = 1$; $\omega_r = 1$, $\hat\omega = 0.05$ and $\gamma = 0.8$. The time scale is in arbitrary units.}  \label{dissipative_loss}
\end{figure}

\noindent The measure of dissipation can be presented by the decrease of the distribution. For this reason the ${\it P}_{d}(t) = \int_{-\infty}^{\infty} \Psi^{\ast}(x,t)\Psi(x,t) \, \dd x$ is also calculated as a function of time. This plot can be seen in Fig \ref{dissipative_loss}. We note that the real part causes only translation (and it results vibration along the axis $x$), but the area below the curve should remain constant. However, the imaginary part of the potential causes the dissipative behavior decreases the value of this integral. 

%\begin{figure}[h!]
%\centerline{
%\includegraphics[width=0.5\columnwidth]{dissipative_loss.eps}}
%\caption{The decrease of ${\it P}_{d}(t) = \int_{-\infty}^{\infty} \Psi^{\ast}(x,t)\Psi(x,t) \, \dd x \leq 1$ in time is showing the dissipation during the process. The out-of-line points come from numerical /computational/ errors. The calculation is evaluated by the parameter set: $m = 1$; $\hslash = 1$; $\omega_r = 1$, $\hat\omega = 0.05$ and $\gamma = 0.8$.}  \label{dissipative_loss}
%\end{figure}

\begin{figure}[h!]
\centerline{
\includegraphics[width=0.5\columnwidth]{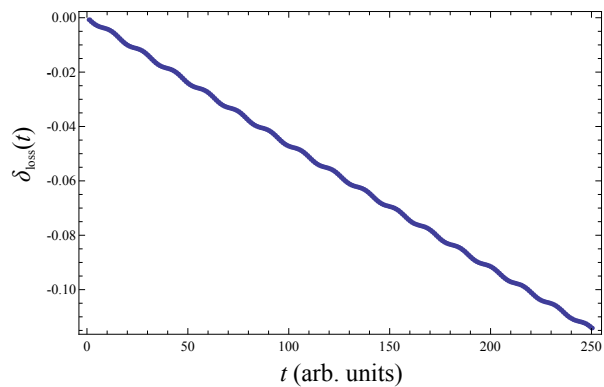}}
\caption{The lost energy $\delta_{loss}(t)$ as a function of time. The calculation is evaluated by the parameter set: $m = 1$; $\hslash = 1$; $\omega_r = 1$, $\hat\omega = 0.05$ and $\gamma = 0.8$. The time scale is in arbitrary units.}  \label{lost_energy}
\end{figure}

\noindent Since, the complex part of the potential generates the dissipation of the oscillator, thus the energy loss $\delta_{\mathrm{loss}}(t)$ (dissipated energy) between the time interval $(0 \rightarrow t)$ relates to the relevant part of the potential $V_{c}(x)$. It can be formulated by the integral 
\begin{equation}
\delta_{\mathrm{loss}}(t) = \int\limits_{0}^{t} \dd t' \, \int\limits_{-\infty}^{\infty} \dd x \, \Psi^{\ast}(x,t') \frac{2}{\hslash}V_{c}(x) \Psi(x,t').
\end{equation}
The lost energy is plotted in  Fig. \ref{lost_energy}.

%\begin{figure}[h!]
%\centerline{
%\includegraphics[width=0.5\columnwidth]{lost_energy.eps}}
%\caption{The lost energy as a function of time. The calculation is evaluated by the parameter set: $m = 1$; $\hslash = 1$; $\omega_r = 1$, $\hat\omega = 0.05$ and $\gamma = 0.8$.}  \label{lost_energy}
%\end{figure}

\section{Conclusions}

We have shown how the kernel of the damped oscillator can be calculated applying the path integral method introducing complex harmonic potentials. Starting from an initial Gaussian wave function the time dependent wave function $\Psi$ of this oscillator could be evaluated. The plotted values of the expression ${\it P}_{d}(t) = \int_{-\infty}^{\infty} \Psi^{\ast}(x,t)\Psi(x,t) \, \dd x \leq 1$ as a function of time clearly shows the dissipation during the damped oscillator motion. The time-dependent energy loss is also calculated.

\section{Supplementary material}

The Hamilton operator of the presented system can be divided into a real and an imaginary part 
\begin{equation}
H = H_r + \cplxi H_c,
\end{equation}
where $H_r$ pertains to 
\begin{equation}
H_r = -\frac{{\hslash}^2}{2m} \nabla^2 + V_r 
\end{equation}
and $H_c < 0$ to the complex part of the potential
\begin{equation}
H_c = V_c . 
\end{equation}
The adjoint operator of $H$ is
\begin{equation}
H^{\dagger} = H_r - \cplxi H_c,
\end{equation}
which means that the Hamilton operator is not Hermitian
\begin{equation}
H \neq H^{\dagger} .
\end{equation}
The time evolution of the system is
\begin{equation}
\cplxi \hslash \frac{\partial\Psi}{\partial t} = H\Psi \,\,\, \longrightarrow \,\,\, \Psi(t) = \expe^{-\frac{\cplxi}{\hslash}\int_{0}^{t} H \, \dd t'} \Psi_0 = U(t,0) \Psi_0,
\end{equation}
where the time evolution  operator 
\begin{equation}
U(t,0) = \expe^{-\frac{\cplxi}{\hslash}\int_{0}^{t} H \, \dd t'} 
\end{equation}
is introduced. For the present case we write
\begin{equation}
U(t,0) = \expe^{-\frac{\cplxi}{\hslash}\int_{0}^{t} H_r \, \dd t' + \frac{1}{\hslash} \int_{0}^{t} H_c \, \dd t'} .
\end{equation}
The $U(t,0)$ completes the time translational symmetry if
\begin{equation}
U^{-1}(t,0) = U^{\dagger}(t,0)
\end{equation}
holds. Now,
\begin{equation}
U^{-1}(t,0) = \expe^{\frac{\cplxi}{\hslash}\int_{0}^{t} H_r \, \dd t' - \frac{1}{\hslash} \int_{0}^{t} H_c \, \dd t'} 
\end{equation}
and
\begin{equation}
U^{\dagger}(t,0) = \expe^{\frac{\cplxi}{\hslash}\int_{0}^{t} H_r \, \dd t' + \frac{1}{\hslash} \int_{0}^{t} H_c \, \dd t'}, 
\end{equation}
from it can be immediately recognized that
\begin{equation}
U^{-1}(t,0) \neq U^{\dagger}(t,0),
\end{equation}
i.e., the time translational symmetry is broken which is expected due to the irreversible behavior. The oscillator energy is not conserved, i.e., the dissipative behavior is involved in the theory in a natural way.

\section{Acknowledgment}

This work is connected to the scientific program of the ''Development of quality-oriented and harmonized R+D+I strategy and functional model at BME'' project. This project is supported by the New Hungary Development Plan (Project ID: T\'AMOP-4.2.1/B-09/1/KMR-2010-0002). This work was also supported by ERC Grant No. ERC-259374-Sylo.

\section*{References}
%\bibliography{markus}

%Figures

\end{document}